\documentclass[11pt]{article}
\usepackage{latexsym,amsmath,amscd,amssymb,graphicx,amsbsy,color,xcolor,amsfonts,amsthm}
\newtheorem{defn}{Definition}

\newtheorem{claim}{Claim}
\begin{document}

\title{Spacetime Transformations from a Uniformly Accelerated Frame\thanks{Supported in part by the German-Israeli Foundation for Scientific Research and Development: GIF No. 1078-107.14/2009}}

\author{Yaakov Friedman and Tzvi Scarr\\
Jerusalem College of Technology\\
Departments of Mathematics and Physics \\
P.O.B. 16031 Jerusalem 91160, Israel\\
e-mail: friedman@jct.ac.il, tzviscarr@gmail.com}

\date{}
\maketitle

\begin{abstract}
We use Generalized Fermi-Walker transport to construct a \emph{one-parameter family} of inertial frames which are instantaneously comoving to a uniformly accelerated observer. We explain the connection between our approach and that of Mashhoon. We show that our solutions of uniformly accelerated motion have constant acceleration in the comoving frame. Assuming the Weak Hypothesis of Locality, we obtain local spacetime transformations from a uniformly accelerated frame $K'$ to an inertial frame $K$. The spacetime transformations between two uniformly accelerated frames with the same acceleration are \emph{Lorentz}. We compute the metric at an arbitrary point of a uniformly accelerated frame.


\end{abstract}


\section{Introduction}\label{Intro}
$\;\;\;$
The accepted physical definition of \emph{uniformly accelerated motion} is motion whose acceleration is \emph{constant in the comoving frame}. This definition is found widely in the literature, as early as \cite{Born2}, again in \cite{LL}, and as recently as \cite{Rohrlich2} and \cite{Lyle}. The best-known example of uniformly accelerated motion is
\emph{one-dimensional hyperbolic motion}. Such motion is exemplified by a particle freely falling in a homogeneous gravitational field. \emph{Fermi-Walker transport} attaches
an instantaneously comoving frame to the particle, and one easily checks that the particle's acceleration is constant in this frame (see \cite{mtw}, pages 166-170, or section
\ref{laexam} below).

In \cite{FS1}, we showed that 1D hyperbolic motion is \emph{not} Lorentz invariant. These motions are, however, contained in a Lorentz-invariant set of motions which we call \emph{translation acceleration}. Moreover, we introduced three new Lorentz-invariant classes of uniformly accelerated motion. For \emph{null} acceleration, the worldline of the motion is \emph{cubic} in the time. \emph{Rotational} acceleration covariantly extends pure rotational motion. \emph{General} acceleration is obtained when the translational component of the acceleration is parallel to the axis of rotation. A review of these explicit solutions appears in section \ref{4types}.

In this paper, we establish that all four types (null, translational, rotational and general) do, in fact, represent uniformly accelerated motion by showing that they have constant acceleration in the instantaneously comoving frame. Fermi-Walker transport will no longer be adequate here to define the comoving frame because we must now deal with \emph{rotating} frames. Instead, we will use \emph{generalized Fermi-Walker transport}. Similar constructions appear in \cite{Mash3} and \cite{mtw}.

Our construction begins in section \ref{defcmf}, where we define the notion of a \emph{one-parameter family} of inertial frames which are instantaneously comoving to a uniformly accelerated observer. As mentioned above, the construction uses Generalized Fermi-Walker transport. This leads us, in section \ref{uaframe}, to the definition of a \emph{uniformly accelerated frame}. Here, we explain the connection between our approach and that of Mashhoon \cite{Mash3}. It is also here that we show that the four types of acceleration
all have constant acceleration in the comoving frame. We also show here that if $K'$ and $K''$ are two uniformly accelerated frames with a common acceleration, then the spacetime transformations between $K'$ and $K''$ are \emph{Lorentz}, despite the fact that neither $K'$ nor $K''$ is inertial.

The main results appear in section \ref{spacetimetrans}. Assuming the Weak Hypothesis of Locality, we obtain local spacetime transformations from a uniformly accelerated frame $K'$ to an inertial frame $K$. We show that these transformations extend the Lorentz transformations between inertial systems. We also compute the metric at an arbitrary point of a uniformly accelerated frame.

Section \ref{examples} is devoted to examples of uniformly accelerated frames and the corresponding spacetime transformations. We summarize our results in section \ref{conc}.

\section{Four Lorentz-invariant Types of Uniformly Accelerated Motion}\label{4types}
$\;\;\;$
In \cite{FS1}, \emph{uniformly accelerated motion} is defined as a motion whose four-velocity $u(\tau)$ in an inertial frame is a
solution to the initial value problem
\begin{equation}\label{uam3}
c\frac{du^{\mu}}{d\tau}=A^{\mu}_{\;\;\nu}u^{\nu}\quad , \quad u(0)=u_0,
\end{equation}
where $A_{\mu\nu}$ is a rank $2$ antisymmetric tensor. Equation (\ref{uam3}) is Lorentz covariant and extends the 3D relativistic dynamics equation $\mathbf{F}=\frac{d\mathbf{p}}{dt}$. The solutions to equation (\ref{uam3}) are divided into four Lorentz-invariant classes: null acceleration, translational acceleration, rotational acceleration, and general acceleration. The translational class is a \emph{covariant extension} of 1D hyperbolic motion and contains the motion of an object in a \emph{homogeneous gravitational field}. In \cite{FS1}, we computed \emph{explicit} worldlines for each of the four types of uniformly accelerated motion. We think of these worldlines as those of a uniformly accelerated \emph{observer}.

Recall that in $1+3$ decomposition of Minkowski space, the acceleration tensor $A$ of equation (\ref{uam3}) has the form
\begin{equation}\label{aab}
A_{\mu\nu}(\mathbf{g},\boldsymbol{\omega})=\left(\begin{array}{cc}0 & \mathbf{g}^T\\ &
\\-\mathbf{g}&-c\mathbf{\Omega}\end{array}\right),
\end{equation}
where $\mathbf{g}$ is a 3D vector with physical dimension of acceleration, $\boldsymbol{\omega}$ is a 3D vector with physical dimension $1/\hbox{time}$, the superscript $T$ denotes matrix transposition, and, for any 3D vector $\boldsymbol{\omega}=(\omega^1,\omega^2,\omega^3)$,
\[ \mathbf{\Omega}= \varepsilon_{ijk}\omega^k, \]
where $\varepsilon_{ijk}$ is the Levi-Civita tensor. The factor $c$ in $A$ provides the necessary physical dimension of acceleration. The 3D vectors $\mathbf{g}$ and $\boldsymbol{\omega}$ are related to the translational acceleration and the angular velocity, respectively, of a uniformly accelerated motion.

We raise and lower indices using the Minkowski metric $\eta_{\mu\nu}=\operatorname{diag}(1,-1,-1,-1)$. Thus, $A_{\mu\nu}=\eta_{\mu\alpha}A^{\alpha}_{\;\;\nu}$, so
\begin{equation}\label{aab2}
A^{\mu}_{\;\;\nu}(\mathbf{g},\boldsymbol{\omega})=\left(\begin{array}{cc}0 & \mathbf{g}^T\\ &
\\\mathbf{g}&c\mathbf{\Omega}\end{array}\right).
\end{equation}

Using the fact that the unique solution to (\ref{uam3}) is given by the exponential function
\begin{equation}\label{exponent solution}
    u(\tau)=\exp (A\tau/c)u_0=\left( \sum_{n=0}^{\infty}\frac{A^n}{n!c^n}\tau ^n\right)u_0\, ,
\end{equation}
we found in \cite{FS1} that the general solution to (\ref{uam3}) is
\begin{equation}\label{gensolnbytype}
u(\tau)=\left\{\begin{array}{l}
  u(0)+Au(0)\tau/c+\frac{1}{2}A^2u(0)\tau^2/c^2,\\
  \;\;\hbox{if }\alpha=0,\beta=0 \hbox{ (null acceleration)}\\
  \;\\
 D_0\cosh(\alpha\tau/c)+D_1\sinh(\alpha\tau/c)+D_2,\\
  \;\;\hbox{if }\alpha>0,\beta=0 \hbox{ (translational acceleration)}\\
 \;\\
 D_0+D_2\cos(\beta\tau/c)+D_3\sin(\beta\tau/c),\\
 \;\;\hbox{if }\alpha=0,\beta>0 \hbox{ (rotational acceleration)} \\
 \;\\
 D_0\cosh(\alpha\tau/c)+D_1\sinh(\alpha\tau/c)\\
+ D_2\cos(\beta\tau/c)+D_3\sin(\beta\tau/c),\\
\;\;\hbox{if }\alpha>0,\beta>0 \hbox{ (general acceleration)}
 \end{array} \right\},
\end{equation}
where $\pm\alpha$ and $\pm i \beta$ are the eigenvalues of $A$, and the $D_\mu$ are appropriate constant four-vectors which depend on $A$ and can be computed explicitly. By integrating $u(\tau)$, one obtains the worldline of a uniformly accelerated observer.

The four classes indicated in (\ref{gensolnbytype}) (null, translational, rotational and general) are \emph{Lorentz-invariant}. The translational class is a covariant extension of 1D hyperbolic motion. The null, rotational, and general classes were previously unknown.

\section{Uniformly Accelerated Frame}\label{ccmf}
$\;\;\;$

In this section, we use Generalized Fermi-Walker transport to define the notion of the \emph{comoving frame of a uniformly accelerated observer}. We then show that in this comoving frame, all of our solutions to equation (\ref{uam3}) have \emph{constant acceleration}. We show that our definition of the comoving frame is equivalent to that of Mashhoon \cite{Mash3}. We also show that if two uniformly accelerated frames have a common acceleration tensor $A$, then the spacetime transformations between them are \emph{Lorentz}, despite the fact that neither frame is inertial.

\subsection{One-Parameter Family of Inertial Frames}\label{defcmf}
$\;\;\;$

First, we define the notion of a one-parameter family of inertial frames which are instantaneously comoving to a uniformly accelerated observer. The coordinates in this family of comoving frames will be used as a bridge between the observer's coordinates and the coordinates in the lab frame $K$. The family of frames is constructed by Generalized Fermi-Walker transport of the initial frame $K_0$ along the worldline of the observer.  In the case of 1D hyperbolic motion, this construction reduces to Fermi-Walker transport \cite{H,Hehl}.

In fact, Fermi-Walker transport may \emph{only} be used in the case of 1D hyperbolic motion. This is because Fermi-Walker transport uses only a part of the Lorentz group - the boosts. This subset of the group, however, is not a \emph{subgroup}, since the combination of two boosts entails a rotation. Generalized Fermi-Walker transport, on the other hand, uses the full homogeneous Lorentz group, and can be used for all four types of uniform acceleration: null, linear, rotational, and general.

The construction of the one-parameter family $\{K_\tau:\tau \ge 0\}$ is according to the following definition.
\begin{defn}\label{cmdefeqn}
Let $\widehat{x}(\tau)$ be the worldline of a uniformly accelerated observer whose motion is determined by the acceleration tensor $A$, the initial four-velocity $u(0)$, and the initial position $\widehat{x}(0)$.

We first define the \emph{initial frame $K_0$}. The origin of $K_0$ at time $\tau=0$ is $\widehat{x}(0)$. For the basis of $K_0$, choose any orthonormal basis $\widehat{\lambda}=\{u(0),\widehat{\lambda}_{(1)},\widehat{\lambda}_{(2)},\widehat{\lambda}_{(3)}\}$.

Next, we define the one-parameter family $\{K_\tau(A,\widehat{x}(0),\widehat{\lambda})\}$ of inertial frames generated by the uniformly accelerated observer. For each $\tau>0$, define $K_{\tau}$ as follows. The origin of $K_{\tau}$ at time $\tau$ is set as $\widehat{x}(\tau)$. The basis of $K_{\tau}$ is defined to be the unique solution
$\lambda(\tau)=\{\lambda_{(\kappa)}(\tau):\kappa=0,1,2,3\}$, to the initial
value problem
\begin{equation}\label{uamlam2}
c\frac{d\lambda_{(\kappa)}^{\mu}}{d\tau}=A^{\mu}_{\;\;\nu}\lambda_{(\kappa)}^{\nu}\quad
, \quad \lambda_{(\kappa)}(0)=\widehat{\lambda}_{(\kappa)}.
\end{equation}
\end{defn}

We remark that the choice of the initial four-velocity $u(0)$ for $\widehat{\lambda}_{(0)}$ is deliberate and required by Generalized Fermi-Walker transport.

\begin{claim}\label{lambda0isu}
For all $\tau$, we have $\lambda_{(0)}(\tau)=u(\tau)$.
\end{claim}
This follows immediately from (\ref{uam3}).

\begin{claim}\label{clambdasoln}
The unique solution to (\ref{uamlam2}) is
\begin{equation}\label{lambdasoln}
\lambda_{(\kappa)}(\tau)=\exp(A\tau/c)\widehat{\lambda}_{(\kappa)}.
\end{equation}
\end{claim}
This follows immediately from  (\ref{exponent solution}).

\begin{claim}\label{ortho}
For all $\tau$, the columns of $\lambda(\tau)$ are an orthonormal basis.
\end{claim}
To prove this claim, it is enough to note that $A$ is antisymmetric. Therefore, $\exp(A\tau/c)$ is an \emph{isometry}.
\vskip0.2cm
Analogously to (\ref{gensolnbytype}), the general solution to (\ref{uamlam2}) is
\begin{equation}\label{gensolncmf}
\!\!\!\!\!\!\!\!\lambda_{(\kappa)}(\tau)=\left\{\begin{array}{l}
  \widehat{\lambda}_{(\kappa)}+A\widehat{\lambda}_{(\kappa)}\tau/c+\frac{1}{2}A^2\widehat{\lambda}_{(\kappa)}\tau^2/c^2,\\
  \;\;\hbox{if }\alpha=0,\beta=0 \hbox{ (null acceleration)}\\
  \;\\
 D_0(\widehat{\lambda}_{(\kappa)})\cosh(\alpha\tau/c)+D_1(\widehat{\lambda}_{(\kappa)})\sinh(\alpha\tau/c)
 +D_2(\widehat{\lambda}_{(\kappa)}),\\
  \;\;\hbox{if }\alpha>0,\beta=0 \hbox{ (linear acceleration)}\\
 \;\\
 D_0(\widehat{\lambda}_{(\kappa)})+D_2(\widehat{\lambda}_{(\kappa)})\cos(\beta\tau/c)
 +D_3(\widehat{\lambda}_{(\kappa)})\sin(\beta\tau/c),\\
 \;\;\hbox{if }\alpha=0,\beta>0 \hbox{ (rotational acceleration)} \\
 \;\\
 D_0(\widehat{\lambda}_{(\kappa)})\cosh(\alpha\tau/c)+D_1(\widehat{\lambda}_{(\kappa)})\sinh(\alpha\tau/c)\\
+ D_2(\widehat{\lambda}_{(\kappa)})\cos(\beta\tau/c)+D_3(\widehat{\lambda}_{(\kappa)})\sin(\beta\tau/c),\\ \;\;\hbox{if }\alpha>0,\beta>0 \hbox{ (general acceleration)}
 \end{array} \right\}.
\end{equation}

\begin{claim}\label{invartype}
For a given $A$, all four solutions $\lambda_{(\kappa)}(\tau), \kappa=0,1,2,3$ are of the same type (null, linear, rotational, or general).
\end{claim}
This claim follows from the fact that the type of acceleration is based solely on the eigenvalues of $A$.

\begin{claim}\label{constantA}
Let $A$ denote the acceleration tensor as computed in the lab frame $K$, and let $\widetilde{A}(\tau)$ denote the tensor as computed in the frame $K_\tau$.
Then $\widetilde{A}(\tau)$ is constant for all $\tau$.
\end{claim}
To prove this claim, first note that $\lambda(\tau)$ is the change of matrix basis from $K$ to $K_\tau$. Hence, using claim \ref{clambdasoln} and the fact that $A$ and $\exp(A\tau/c)$ commute, we have
\[ \widetilde{A}(\tau)=\lambda(\tau)^{-1}A\lambda(\tau)=(\exp(A\tau/c)\widehat{\lambda})^{-1}A\exp(A\tau/c)\widehat{\lambda}\]
\begin{equation}\label{alla}
=  \widehat{\lambda}^{-1}\exp(A\tau/c)^{-1}A\exp(A\tau/c)\widehat{\lambda}= \widehat{\lambda}^{-1}\exp(A\tau/c)^{-1}\exp(A\tau/c)A\widehat{\lambda}= \widehat{\lambda}^{-1} A \widehat{\lambda}=\widetilde{A}(0).
\end{equation}

\begin{claim}\label{lambdaAisAlambda}
For all $\tau$, we have $\lambda(\tau)\widetilde{A}(\tau)=A\lambda(\tau)$.
\end{claim}
This follows from the first equality in (\ref{alla}).

\subsection{Uniformly Accelerated Frame}\label{uaframe}
$\;$

Two frames are said to be \emph{comoving} at time $\tau$ if at this time, the origins and the axes of the two frames coincide, and they have the same four-velocity.

We now define the notion of a \emph{uniformly accelerated frame}.

\begin{defn}\label{defuasys}
A frame $K'$ is \emph{uniformly accelerated} if there exists a one-parameter family $\{K_\tau(A,\widehat{x}(0),\widehat{\lambda})\}$ of inertial frames generated by a uniformly accelerated observer such that at every time $\tau$, the frame $K_\tau$ is comoving to $K'$.
\end{defn}
In light of this definition, we may regard our uniformly accelerated observer as positioned at the spatial origin of a uniformly accelerated frame. This approach is motivated by the following statement of Brillouin \cite{Brill}: a frame of reference is a ``heavy
laboratory, built on a rigid body of tremendous mass, as compared to the masses
in motion."

Our construction of a uniformly accelerated frame should be contrasted with Mashhoon's approach \cite{Mash3}, which is well suited to curved spacetime, or a manifold setting.
There, the orthonormal basis is defined by
\begin{equation}\label{Mashhoondeflam2}
 c\frac{d\lambda_{(\kappa)}^{\mu}(\tau)}{d\tau}=\widetilde{A}^{(\nu)}_{\;\;(\kappa)}\lambda_{(\nu)}^{\mu}(\tau),
\end{equation}
where $\widetilde{A}$ is a constant antisymmetric tensor. Notice that the derivative of each of Mashhoon's basis vectors depends on \emph{all} of the basis vectors, whereas the derivative of each of our basis vectors depends only on its own components. In particular, Mashhoon's observer's four-acceleration depends on both his four-velocity $\lambda_{(0)}$ \emph{and} on the spatial vectors of his basis, while our observers's four-acceleration depends \emph{only} on his four-velocity. This seems to be the more natural physical model: is there any \emph{a priori} reason why the four-acceleration of the observer should depend on his \emph{spatial} basis? We show now, however, that the two approaches are, in fact, equivalent.

The two approaches are equivalent if we identify Mashhoon's tensor $\widetilde{A}$ with our own tensor $\widetilde{A}(0)$: $\widetilde{A}=\widetilde{A}(\tau)=\widetilde{A}(0)$.
Then, by equation (\ref{uamlam2}) and claim \ref{lambdaAisAlambda}, we have
\[ c\frac{d\lambda_{(\kappa)}^{\mu}(\tau)}{d\tau}=A^{\mu}_{\;\;\nu}\lambda_{(\kappa)}^{\nu}(\tau)
=\lambda_{(\nu)}^{\mu}(\tau)\widetilde{A}^{(\nu)}_{\;\;(\kappa)} , \]
which is (\ref{Mashhoondeflam2}).

We now show that all of our solutions of equation (\ref{uam3}) have constant acceleration in the comoving frame. Let $A$ be as in (\ref{aab}). Denote $\tilde{u}=(1,0,0,0)^T$. By claim 1 and 2 we have
\begin{equation}\label{aisconstgenproof}
a(\tau)=Au(\tau)=A\lambda_{(0)}(\tau)=A\exp(A\tau/c)\hat{\lambda}_{(0)}=A\exp(A\tau/c)\hat{\lambda}\tilde{u}
\end{equation}\[=\exp(A\tau/c)\hat{\lambda}\hat{\lambda}^{-1}A\hat{\lambda}\tilde{u}=\lambda(\tau)\widetilde{A}\tilde{u}=
\lambda_{(i)}(\tau)\widetilde{\mathbf{g}}^{(i)}.\]

Thus, the acceleration of the observer in the comoving frame is constant and equals $\widetilde{\mathbf{g}}$ from the decomposition (\ref{aab}) for $\widetilde{A}$.
\vskip0.3cm

We end this section by showing that if $K'$ and $K''$ are two uniformly accelerated frames with a common acceleration tensor $A$, then the spacetime transformations between $K'$ and $K''$ are \emph{Lorentz}, despite the fact that neither $K'$ nor $K''$ is inertial. Let $K'$ and $K''$ be two uniformly accelerated frames with a common acceleration tensor $A$. Without loss of generality, let the lab frame $K$ be the initial comoving frame $K_0$ of $K'$, so that the initial orthonormal basis of $K'$ is the identity $I$. Let $\widehat{\lambda}$ be the initial orthonormal basis of $K''$. Then the basis of $K'$ at time $\tau$ is $\lambda'(\tau)=\exp(A\tau/c)$, while the basis of $K''$ at time $\tau$ is $\lambda''(\tau)=\exp(A\tau/c)\widehat{\lambda}=\lambda'(\tau)\widehat{\lambda}$. Thus, the change of basis from $K'$ to $K''$ is accomplished by the Lorentz transformation with matrix representation $\widehat{\lambda}$. This implies, in particular, that there is a Lorentz transformation from a lab frame on Earth to an airplane flying at constant velocity, since they are both subject to the same gravitational field.

\section{Spacetime Transformations from a uniformly accelerated frame to the lab frame}\label{spacetimetrans}

$\;\;\;$
In this section, we construct the spacetime transformations from a uniformly accelerated frame $K'$ to the lab frame $K$. This will be done in two steps.
\vskip0.4cm\noindent
\large{\textbf{Step 1: From $K_\tau$ to $K$}}
\vskip0.2cm\noindent
First, we will derive the spacetime transformations from $K_\tau$ to $K$. The idea here is as follows. Fix an event $X$ with  coordinates $x^\mu$ in $K$. Find the time $\tau$ for which $\widehat{x}(\tau)$ is simultaneous to $X$ in the comoving frame $K_\tau$. Define the $0$-coordinate in $K_\tau$ to be $y^{(0)}=c\tau$. Use the basis $\lambda(\tau)$ of $K_\tau$ to write the relative spatial displacement of the event $X$ with respect to the observer as $y^i\lambda_{(i)}(\tau),i=1,2,3$. The spacetime transformation from $K_\tau$ to $K$ is then defined to be
\begin{equation}\label{sttransexp}
x^\mu=\widehat{x}^\mu(\tau)+y^{(i)}\lambda_{(i)}^\mu(\tau).
\end{equation}
\begin{figure}[h!]
  \centering
\scalebox{0.8}{\includegraphics{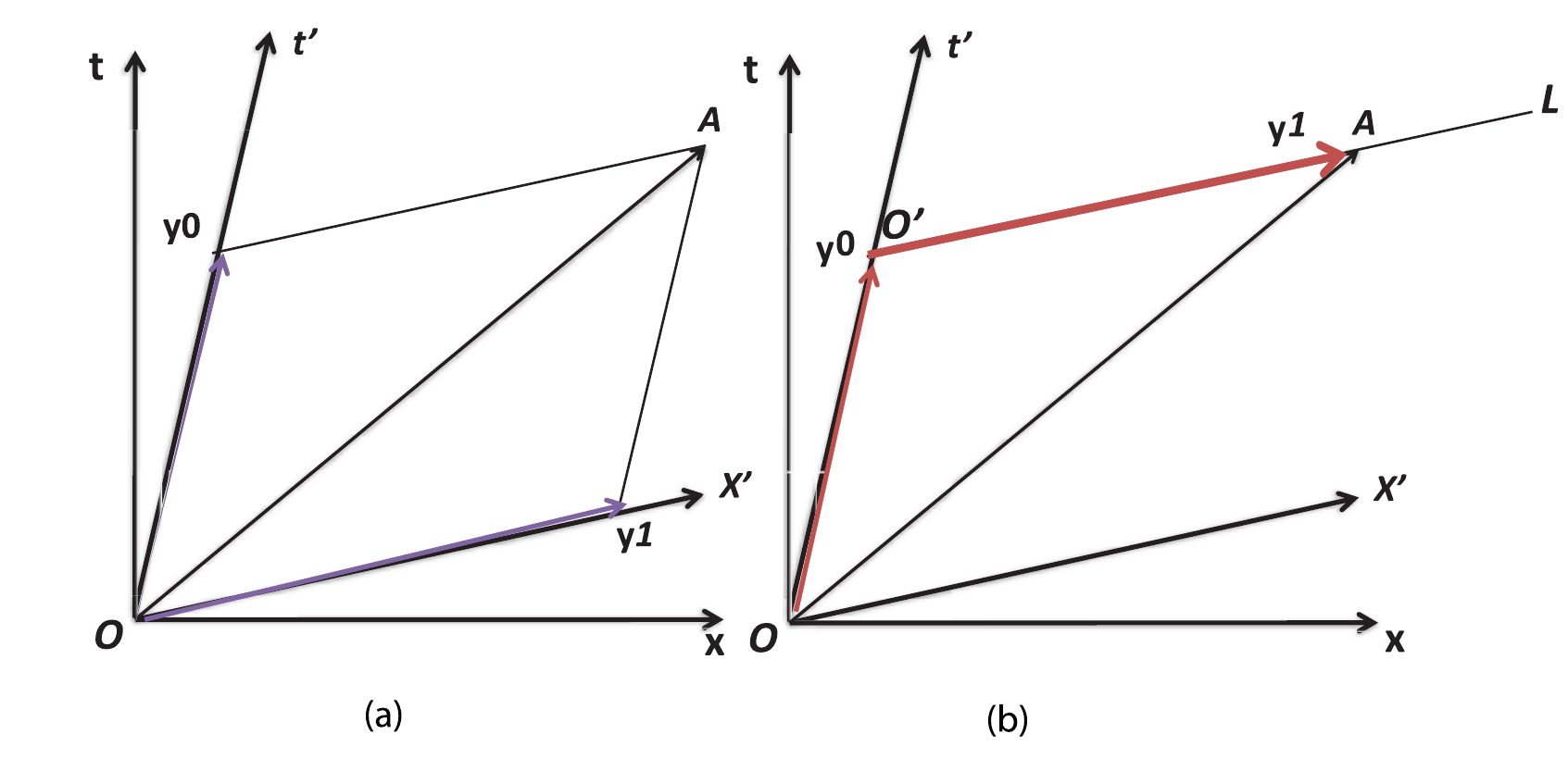}}
  \caption{Lorentz transformation\;-\;two perspectives. (a) The usual perspective (b) Our approach }
\end{figure}
Transformations of the form (\ref{sttransexp}) have a natural physical interpretation and were also used in \cite{Mash3}. Moreover, they extend the \emph{Lorentz transformations}.

To see this, let $K'$ be an \emph{inertial} frame (simply set $A=0$). We will show that the Lorentz transformations $K'\rightarrow K$ can be written as in (\ref{sttransexp}).
Suppose that $K'$ moves with 3D velocity $\mathbf{v}=(v,0,0)$ with respect to $K$. Assume, as usual,
that the observer located at the spatial origin of $K'$ was at the origin of $K$ at time $t=0$.  Let  $x^\mu=(x^0=ct,x^i)$ denote the coordinates of an event in $K$, and let $y^{(\mu)}$ denote the event's coordinates in $K'$. In $K$, the observer has constant four-velocity $\widehat{u}=\gamma(1,v/c,0,0)$, and the observer's worldline in $K$ is $\widehat{x}(\tau)=c\tau\widehat{u}=\widehat{u}y^{(0)}$. In this case, the comoving frame of $K'$ is
\begin{equation}\label{comovIner}
 \lambda_{(0)}=\gamma(1,v/c,0,0), \lambda_{(1)}=\gamma(v/c,1,0,0),\lambda_{(2)}=(0,0,1,0),\lambda_{(3)}=(0,0,0,1).
\end{equation}

The Lorentz transformations $K' \rightarrow K$ are usually written as
\[ x=(ct,x^1,x^2,x^3)=(\gamma(c\tau+vy^{(1)}/c),\gamma(v\tau+y^{(1)}),y^{(2)},y^{(3)}).\]
In this form, the transformations correspond to Figure 1(a), in which the event $A$ is written as a linear combination of unit vectors along the $x'$ and $t'$ axes.

However, we can write these transformations equivalently as
\[x=c\tau \gamma(1,v/c,0,0)+y^{(1)}\gamma(v/c,1,0,0) +y^{(2)}(0,0,1,0)+y^{(3)}(0,0,0,1), \]
which is exactly
\begin{equation}\label{lta0}
x=\widehat{x}(\tau)+y^{(i)}\lambda_{(i)}.
\end{equation}
In this form, the transformations correspond to Figure 1(b), in which the event $A$ is written as the vector sum of the worldline of the observer located at the origin of $K'$ and the event's spatial coordinates in this observer's comoving frame.

It is worthwhile noting the properties of the transformations (\ref{lta0}) when $K'$ is inertial ($A=0$). First of all, only when $K'$ is inertial are the transformations (\ref{lta0}) linear, since only in this case does the observer's position depend linearly on $y^{(0)}$.
Next, note that when $K'$ is inertial, the transformations (\ref{lta0}) are well defined on all of Minkowski space. For each value of $\tau$, let $X_{\tau}$ be the
3D spacelike hyperplane consisting of all events simultaneous (in $K_{\tau}$) to $\widehat{x}(\tau)$. These hyperplanes are parallel and, therefore, pairwise disjoint. Now, let $X$ be an event with coordinates $y^{(0)},y^{(1)},y^{(2)},y^{(3)}$ in $K'$. This event is simultaneous to the event $\widehat{x}(\tau_0)=(y^{(0)},0,0,0)$, which corresponds to the observer at time $\tau_0=y^{(0)}/c$.  The vector $X-\widehat{x}(\tau_0)$ belongs to the hyperplane
$X_{\tau_0}$ and may therefore be decomposed as $X-\widehat{x}(\tau_0)=y^{(i)}\lambda_{(i)}(\tau_0)$ in $K_{\tau_0}$. Since the $X_\tau$ are pairwise disjoint, the vector $X-\widehat{x}(\tau_0)$ does not belong to any other $X_\tau$. Hence, the transformations are well defined everywhere.

Returning to the general case ($A\neq 0$), we are now ready to show that the spacetime transformations from $K_\tau$ to $K$ have the form (\ref{lta0}).  Let $K'$ be the uniformly accelerated frame determined by $A$, $\widehat{x}(0)$, and $\widehat{\lambda}$. The worldline $\widehat{x}(\tau)$ of the observer is obtained by integrating his four-velocity $u(\tau)$, and the comoving frame matrix $\lambda(\tau)$ is given by (\ref{gensolncmf}). In order to use (\ref{lta0}), it remains only to establish well-defined spatial coordinates $y^{(\mu)}$ in $K_\tau$.

Since $K'$ is accelerated, the hyperplanes $X_{\tau}$ are no longer pairwise disjoint. Nevertheless, since $X_{\tau}$
is \emph{perpendicular} to $u(\tau)$, there exist a neighborhood of $\tau$ and a spatial neighborhood of the observer in which the $X_{\tau}$ are
pairwise disjoint. Thus, spacetime can be \textit{locally} split into disjoint 3D spacelike hyperplanes. This insures that, at least locally, the same event does not occur at two different times. Hence, within the locality restriction, we may uniquely define coordinates for the observer. This implies that, at least \emph{locally}, the spacetime transformations
from $K_\tau$ to $K$ are given by (\ref{lta0}).

A similar construction can be found in \cite{al1}, in which the authors use \emph{radar 4-coordinates}.
\vskip0.4cm\noindent
\large{\textbf{Step 2: From $K'$ to $K_\tau$}}
\vskip0.2cm\noindent
At this point, we invoke a weaker form of the Hypothesis of Locality introduced by
Mashhoon \cite{Mash1,Mash2}. This \emph{Weak Hypothesis of
Locality} is an extension of the Clock Hypothesis.
\vskip0.2cm \textbf{The Weak Hypothesis of
Locality}\quad \emph{Let $K'$ be a uniformly accelerated frame, with an accelerated
observer with worldline $\widehat{x}(\tau)$. For any time $\tau_0$, the rates of the clock of
the accelerated observer and the clock at the origin of the comoving frame $K_{\tau_0}$ are the
same, and, for events simultaneous to $\widehat{x}(\tau_0)$ in the comoving frame $K_{\tau_0}$, the comoving and the accelerated observers measure the same spatial components.}
\vskip0.2cm
Consider an event with $K$ coordinates $x^{\mu}$. By step 1, we have, for a unique $\tau_0$, $x=\widehat{x}(\tau_0)+y^{(i)}\lambda_{(i)}(\tau_0)$. Hence, the $K_{\tau_0}$
coordinates of the event $x$ are $y^{(0)}=c\tau_0$ and $y^{(i)}$. Since $x$ and $\widehat{x}(\tau_0)$ are
simultaneous in $K_{\tau_0}$, the Weak Hypothesis of Locality implies that the spatial
coordinates $y^{(i)}$ coincide with the spatial coordinates in $K'$. We have thus proven the following:
\vskip0.4cm
Let $K'$ be a uniformly accelerated frame attached to an observer with worldline $\widehat{x}(\tau)$. Let $\{K_\tau(A,\widehat{x}(0),\widehat{\lambda})\}$ be the corresponding one-parameter family of inertial frames. Then the
\textit{the spacetime transformations from $K'$ to $K$} are
\begin{equation}\label{xy}
x=\widehat{x}(\tau)+y^{(i)}\lambda_{(i)}(\tau), \;\;\mbox{ with}\;\; \tau=y^{(0)}/c.
\end{equation}
\vskip0.4cm
Unless specifically mentioned otherwise, we will always choose the lab frame $K$ to be the initial comoving frame $K_0$. This implies that $\widehat{\lambda}=I$ and $A=\widetilde{A}$.

We end this section by calculating the metric at the point $y$ of $K'$. First, we calculate the
differential of the transformation (\ref{xy}). Differentiating (\ref{xy}), we have
\[dx= \lambda_{(0)}(\tau)
dy^{(0)}+\lambda_{(i)}(\tau)dy^{(i)}+
y^{(i)}\frac{1}{c}\frac{d\lambda_{(i)}}{d\tau}dy^{(0)}.\]
Define $\bar{y}=(0,\mathbf{y})$. Using (\ref{Mashhoondeflam2}) (but writing $A$ for $\widetilde{A}$, as is our convention), this becomes
\begin{equation}\label{differential gen2}
dx=\lambda_{(0)}(\tau)
dy^{(0)}+\lambda_{(i)}(\tau)dy^{(i)}+c^{-2}(A\bar{y})^{(\nu)}\lambda_{(\nu)}(\tau)dy^{(0)}\,.
\end{equation}
Finally, since
\begin{equation}\label{Aactson4vector}
A\bar{y}=(\mathbf{g}\cdot \mathbf{y},\mathbf{y}\times c \boldsymbol{\omega}),
\end{equation}
we obtain
\begin{equation}\label{differential gen1}
dx=\left(\left(1+\frac{\mathbf{g}\cdot\mathbf{y}}{c^2}\right)\lambda_{(0)}+c^{-1}(\mathbf{y}\times
\boldsymbol{\omega})^{(i)}\lambda_{(i)}\right)dy^{(0)}+\lambda_{(j)}dy^{(j)}.
\end{equation}
Therefore, the metric at the point $\bar{y}$ is

\[ s^2=dx^2=\left( \left(1+\frac{\mathbf{g}\cdot\mathbf{y}}{c^2}\right)^2-c^{-2}(\mathbf{y}\times
\boldsymbol{\omega})^2\right)(dy^{(0)})^2 \]
\begin{equation}\label{metricaty}
+\frac{2}{c}(\mathbf{y}\times
\boldsymbol{\omega})_{(i)}dy^{(0)}dy^{(i)}+\delta_{jk}dy^{(j)}dy^{(k)}.
\end{equation}
This formula was also obtained by Mashhoon \cite{Mash3}. We point out that the metric is dependent only on the \emph{position} in the accelerated frame and not on \emph{time}.

\section{Examples of Spacetime Transformations from a Uniformly Accelerated frame}\label{examples}

$\;\;\;$
In this section, we consider examples of uniformly accelerated frames and the corresponding spacetime transformations. \vskip0.5cm\noindent

\subsection{Null Acceleration\quad ($\alpha=0,\beta=0$)}

Since, in this case, $|\mathbf{g}|=|c\boldsymbol{\omega}|$ and $\mathbf{g} \cdot \boldsymbol{\omega}=0$, we may choose $\mathbf{g}=(g,0,0)$ and
$c\boldsymbol{\omega}=(0,0,g)$.
From (\ref{aab}), we have
\begin{equation}\label{Atype0ex1}
A^{\mu}_{\;\;\nu}=\left(\begin{array}{cccc}0 & g & 0 &0\\g&0&g&0 \\
0&-g&0&0\\0&0&0&0\end{array}\right).
\end{equation}
Then
\begin{equation}\label{Asqtype0ex1}
A^2=\left(\begin{array}{cccc}g^2 & 0 & g^2 & 0\\0&0&0&0 \\
-g^2&0&-g^2&0\\0 & 0 & 0 &0\end{array}\right).
\end{equation}
Thus, from (\ref{gensolncmf}), we have
\begin{equation}\label{lambdatype0ex1}
\lambda(\tau)=I+A\tau/c+\frac{1}{2}A^2\tau^2/c^2=\left(\begin{array}{cccc}
1+\frac{g^2\tau^2}{2c^2} & g\tau/c & \frac{g^2\tau^2}{2c^2} & 0 \\ g\tau/c & 1 & g\tau/c & 0 \\
-\frac{g^2\tau^2}{2c^2} & -g\tau/c & 1-\frac{g^2\tau^2}{2c^2} & 0 \\ 0 & 0 & 0 & 1
\end{array}\right).
\end{equation}
The observer's four-velocity is, therefore,
\begin{equation}\label{lambda0type0ex1}
u(\tau)=\lambda_{(0)}(\tau)=\left(1+\frac{g^2\tau^2}{2c^2},g\tau/c,-\frac{g^2\tau^2}{2c^2},0 \right).
\end{equation}
His four-acceleration is
\begin{equation}\label{acceltype0ex1}
a(\tau)= \left(\frac{g^2\tau}{c},g,-\frac{g^2\tau}{c},0 \right)=g\lambda_{(1)}(\tau),
\end{equation}
which shows that the acceleration is constant in the comoving frame.

Integrating (\ref{lambda0type0ex1}), we have
\[ \widehat{x}(\tau)=    \left( c\tau+\frac{g^2\tau^3}{6c},\frac{g\tau^2}{2},-\frac{g^2\tau^3}{6c},0\right). \]
Using (\ref{lambdatype0ex1}) and $y^{(0)}=c\tau$, the spacetime transformations (\ref{xy}) are
\begin{equation}\label{stttype0}
\left(\begin{array}{c}x^0 \\  \; \\ x^1 \\ \; \\ x^2 \\\; \\ x^3 \end{array}\right)
 =\left( \begin{array}{c}c\tau+ \frac{g^2\tau^3}{6c}+y^{(1)}g\tau/c +y^{(2)}\frac{g^2\tau^2}{2c^2} \\ \; \\ \frac{g\tau^2}{2}+y^{(1)}+y^{(2)}g\tau/c \\ \; \\
  -\frac{g^2\tau^3}{6c}- y^{(1)}g\tau/c+y^{(2)}-y^{(2)}\frac{g^2\tau^2}{2c^2} \\ \; \\ y^{(3)} \end{array} \right).
\end{equation}

\subsection{Linear Acceleration\quad ($\alpha>0,\beta=0$)}\label{laexam}

Without loss of generality, we may choose
\begin{equation}\label{Atype1ex1}
A= \left(\begin{array}{cccc}0 & g & 0 &0\\g&0&c\omega & 0 \\
0&-c\omega&0&0\\0&0&0&0\end{array}\right),
\end{equation}
where $g>c\omega>0$. In order to simplify the calculation of the exponent of $A$, we perform a Lorentz boost
\begin{equation}\label{lineardriftboost}
 B= \left(\begin{array}{cccc}  g/\alpha & 0 & -c\omega/\alpha& 0  \\
0 & 1 & 0 & 0 \\
-c\omega/\alpha & 0 & g/\alpha & 0 \\
 0 & 0 & 0 & 1
 \end{array}\right)
\end{equation}
to the drift frame corresponding to the velocity
\begin{equation}\label{driftvelocitylin}
\mathbf{v} =(c^2/g,0,0)\times (0,0,\omega).
\end{equation}
Since $g>c\omega>0$, we have $|\mathbf{v}|\le c$.

In the drift frame, the acceleration tensor $A$ becomes
\[ A_{dr}=B^{-1}AB=\left(\begin{array}{cccc}  0 & \alpha & 0 & 0  \\
\alpha & 0 & 0 & 0 \\
0 & 0 & 0 & 0\\
 0 & 0 & 0 & 0
 \end{array}\right),\]
and leads to 1D hyperbolic motion.
Hence,
\[ \lambda(\tau)=\exp(A\tau/c)=B\exp(A_{dr}\tau/c)B^{-1}\]
\begin{equation}\label{lambdatype1ex1}
= \left(\begin{array}{cccc}\frac{g^2}{\alpha^2}\left(\cosh\frac{\alpha\tau}{c}-1\right)+1 &
\frac{g}{\alpha}\sinh\frac{\alpha\tau}{c}
 & \frac{gc\omega}{\alpha^2}\left(\cosh\frac{\alpha\tau}{c}-1\right) & 0  \\ \frac{g}{\alpha}\sinh\frac{\alpha\tau}{c}&\cosh\frac{\alpha\tau}{c}&\frac{c\omega}{\alpha}\sinh\frac{\alpha\tau}{c}&
 0\\
\frac{-gc\omega}{\alpha^2}\left(\cosh\frac{\alpha\tau}{c}-1\right)&\frac{-c\omega}{\alpha}\sinh\frac{\alpha\tau}{c}&
\frac{-c^2\omega^2}{\alpha^2}\left(\cosh\frac{\alpha\tau}{c}-1\right)+1&0\\
0&0&0&1\end{array}\right).
\end{equation}
If $\omega=0$, we recover the usual hyperbolic motion of a frame. Thus, the previous formula is a covariant extension of hyperbolic motion.

From the first column of (\ref{lambdatype1ex1}), the observer's four-velocity is
\begin{equation}\label{4veltype1ex1}
u(\tau)= (\frac{g^2}{\alpha^2}\left(\cosh\frac{\alpha\tau}{c}-1\right)+1,\frac{g}{\alpha}\sinh\frac{\alpha\tau}{c},\frac{-gc\omega}{\alpha^2}\left(\cosh\frac{\alpha\tau}{c}-1\right),0)  .
\end{equation}
Hence, the observer's four-acceleration is
\begin{equation}\label{acceltype1ex1}
a(\tau)=  \left(\frac{g^2}{\alpha}\sinh\frac{\alpha\tau}{c},g\cosh\frac{\alpha\tau}{c},\frac{-gc\omega}{\alpha}\sinh\frac{\alpha\tau}{c},0\right)
=g\lambda_{(1)}(\tau),
\end{equation}
which shows that the acceleration is constant in the comoving frame.

Note that our definition of linear acceleration is more general than the usual $\frac{d\mathbf{u}}{dt}  =  \mathbf{g}$. From formula (\ref{4veltype1ex1}), we have
\[ \mathbf{u}=\left(\frac{cg}{\alpha}\sinh\frac{\alpha\tau}{c},
\frac{-gc^2\omega}{\alpha^2}\left(\cosh\frac{\alpha\tau}{c}-1\right),0\right).\]
Since $\frac{d\tau}{dt}=\gamma^{-1}$, and $\gamma$ is the zero component of $u(\tau)$, we have
\[ \frac{d\mathbf{u}}{dt}=\frac{d\mathbf{u}}{d\tau}\frac{d\tau}{dt}=\frac{\left(g\cosh\frac{\alpha\tau}{c},
 \frac{-gc\omega}{\alpha} \sinh\frac{\alpha\tau}{c} ,0 \right)}{\frac{g^2}{\alpha^2}\left(\cosh\frac{\alpha\tau}{c}-1\right)+1},\]
which is not constant unless $\omega=0$. This provides a proof of the fact mentioned in part I that the equation $\frac{d\mathbf{u}}{dt}  =  \mathbf{g}$ is limited to the particular case $\omega=0$.

Integrating (\ref{4veltype1ex1}), we have
\[ \widehat{x}(\tau)=    \left( \frac{c^2}{\alpha^2}\left(\frac{g^2}{\alpha}\sinh\frac{\alpha\tau}{c}+c\omega\tau\right),\frac{c^2g}{\alpha^2}
\left(\cosh\frac{\alpha\tau}{c}-1\right),\frac{-c^2g\omega}{\alpha^2}\left(\frac{c}{\alpha}\sinh\frac{\alpha\tau}{c}-\tau\right),
0\right).\]
Using (\ref{lambdatype1ex1}) and $y^{(0)}=c\tau$, the spacetime transformations (\ref{xy}) are
\begin{equation}\label{stttype1}
\left(\begin{array}{c}x^0 \\  \; \\ \; \\ x^1 \\ \; \\ \; \\ x^2 \\ \; \\ \; \\ x^3 \end{array}\right)
 =\left( \begin{array}{c}\frac{c^2}{\alpha^2}\left(\frac{g^2}{\alpha}\sinh\frac{\alpha\tau}{c}+c\omega\tau\right)+y^{(1)}\frac{g}{\alpha}\sinh\frac{\alpha\tau}{c}\\
  \;\;\;\;\;\;\;\;+y^{(2)}\frac{cg\omega}{\alpha^2}\left(\cosh\frac{\alpha\tau}{c}-1\right)\\ \; \\ \frac{c^2g}{\alpha^2}\left(\cosh\frac{\alpha\tau}{c}-1\right)
  +y^{(1)}\cosh\frac{\alpha\tau}{c}\\ \;\;\;\;\;\;\;\;+y^{(2)}\frac{c\omega}{\alpha}\sinh\frac{\alpha\tau}{c}  \\ \; \\ \frac{-c^2g\omega}{\alpha^2}\left(\frac{c}{\alpha}\sinh\frac{\alpha\tau}{c}-\tau\right)-y^{(1)}\frac{c\omega}{\alpha}\sinh\frac{\alpha\tau}{c}\\
  \;\;\;\;\;\;\;\;-y^{(2)}\frac{c^2\omega^2}{\alpha^2}\left(\cosh\frac{\alpha\tau}{c}-1\right)+y^{(2)}
    \\ \; \\ y^{(3)}   \end{array} \right).
\end{equation}
\subsection{Rotational Acceleration\quad ($\alpha=0,\beta>0$)}

Without loss of generality, we may choose
\begin{equation}\label{Atype2ex1}
A= \left(\begin{array}{cccc}0 & g & 0 &0\\g&0&c\omega&0\\
0&-c\omega&0&0\\0&0&0&0\end{array}\right),
\end{equation}
where $c\omega>g>0$. In order to simplify the calculation of the exponent of $A$, we perform a Lorentz boost
\[B= \left(\begin{array}{cccc}  c\omega/\beta & 0 & -g/\beta& 0  \\
0 & 1 & 0 & 0 \\
-g/\beta & 0 & c\omega/\beta & 0 \\
 0 & 0 & 0 & 1
 \end{array}\right)\]
to the drift frame corresponding to the velocity
\begin{equation}\label{driftvelocityrot}
\mathbf{v} =(g,0,0)\times (0,0,1/\omega).
\end{equation}
Since $c\omega>g>0$, we have $|\mathbf{v}|\le c$.

In the drift frame, the acceleration tensor $A$ becomes
\[ A_{dr}=B^{-1}AB=\left(\begin{array}{cccc}  0 & 0 & 0 & 0  \\
0 & 0 & \beta & 0 \\
0 & -\beta & 0 & 0\\
 0 & 0 & 0 & 0
 \end{array}\right),\]
and leads to pure rotational motion.
Hence,
\[ \lambda(\tau)=\exp(A\tau/c)=B\exp(A_{dr}\tau/c)B^{-1}\]

\begin{equation}\label{lambdatype2ex1}
= \left(\begin{array}{cccc}\frac{g^2}{\beta^2}(1-\cos\frac{\beta\tau}{c})+1 & \frac{g}{\beta}\sin\frac{\beta\tau}{c} & \frac{gc\omega}{\beta^2}(1-\cos\frac{\beta\tau}{c}) & 0 \\ \frac{g}{\beta}\sin\frac{\beta\tau}{c}&\cos\frac{\beta\tau}{c}&\frac{c\omega}{\beta}\sin\frac{\beta\tau}{c}&0\\ \frac{-gc\omega}{\beta^2}(1-\cos\frac{\beta\tau}{c})&-\frac{c\omega}{\beta}\sin\frac{\beta\tau}{c}&\frac{-c^2\omega^2}
{\beta^2}(1-\cos\frac{\beta\tau}{c})+1&0\\0&0&0&1\end{array}\right).
\end{equation}
If $g=0$, we recover the usual rotation of the basis about the $z$ axis. Thus, the previous formula is a covariant extension of rotational motion.

From the first column of (\ref{lambdatype2ex1}), the observer's four-velocity is
\begin{equation}\label{4veltype2ex1}
u(\tau)=\left(\frac{g^2}{\beta^2}(1-\cos\frac{\beta\tau}{c})+1,
 \frac{g}{\beta}\sin\frac{\beta\tau}{c},\frac{-gc\omega}{\beta^2}(1-\cos\frac{\beta\tau}{c}),0 \right).
\end{equation}
Hence, the observer's four-acceleration is
\begin{equation}\label{acceltype2ex1}
a(\tau)=  \left(\frac{g^2}{\beta}\sin\frac{\beta\tau}{c},g\cos\frac{\beta\tau}{c},\frac{-gc\omega}{\beta}\sin\frac{\beta\tau}{c},0\right)
=g\lambda_{(1)}(\tau),
\end{equation}
which shows that the acceleration is constant in the comoving frame.

Integrating (\ref{4veltype2ex1}), we have
\[ \widehat{x}(\tau)= \left( \frac{-c^2g^2}{\beta^3}\sin\frac{\beta\tau}{c}+\left(\frac{g^2}{\beta^2}+1\right)c\tau,
\frac{c^2g}{\beta^2}\left(\cos\frac{\beta\tau}{c}-1\right), \frac{c^3g\omega}{\beta^3}\sin\frac{\beta\tau}{c}-\frac{c^2g\omega}{\beta^2}\tau,0
\right). \]
Using (\ref{lambdatype2ex1}) and $y^{(0)}=c\tau$, the spacetime transformations (\ref{xy}) are
\begin{equation}\label{stttype2}
\left(\begin{array}{c}x^0 \\  \; \\ \; \\ x^1 \\ \; \\ \; \\ x^2 \\ \; \\ \; \\ x^3 \end{array}\right)
 =\left( \begin{array}{c}\frac{-c^2g^2}{\beta^3}\sin\frac{\beta\tau}{c}+\left(\frac{g^2}{\beta^2}+1\right)c\tau-\frac{gy^{(1)}}{\beta}\sin\frac{\beta\tau}{c}\\
+ \frac{y^{(2)}cg\omega}{\beta^2}(1-\cos\frac{\beta\tau}{c})\\ \; \\ \frac{c^2g}{\beta^2}\left(\cos\frac{\beta\tau}{c}-1\right)
+y^{(1)}\cos\frac{\beta\tau}{c}\\-\frac{y^{(2)}c\omega}{\beta}\sin\frac{\beta\tau}{c} \\ \; \\ \frac{c^3g\omega}{\beta^3}\sin\frac{\beta\tau}{c}-\frac{c^2g\omega}{\beta^2}\tau
 + \frac{y^{(1)}c\omega}{\beta}\sin\frac{\beta\tau}{c}\\ - \frac{y^{(2)}c^2\omega^2}{\beta^2}(1-\cos\frac{\beta\tau}{c})+ y^{(2)}\\ \; \\ y^{(3)}   \end{array} \right).
\end{equation}

\section{Summary}\label{conc}
$\;\;\;$

Definition \ref{cmdefeqn} introduces a new method of constructing a family of inertial frames which are instantaneously comoving to a uniformly accelerated observer. Our construction uses generalized Fermi-Walker transport, and we have shown that our approach is equivalent to that of Mashhoon (equation (\ref{Mashhoondeflam2})). Thus, we may use the two approaches interchangeably.  Mashhoon's approach is better suited to curved spacetime, that is, a manifold setting. Our approach, on the other hand, leads to a \emph{decoupled} system of differential equations, and is, therefore, easier to solve for explicit solutions.

Moreover, all of our solutions (\ref{gensolnbytype}) for uniformly accelerated motion have \emph{constant acceleration in the comoving frame} (see equation (\ref{aisconstgenproof})). In fact, the value of this constant acceleration is $\mathbf{g}$, the linear acceleration component of the acceleration tensor $A$.

We have also shown at the end of section \ref{ccmf} that the spacetime transformations between two frames $K'$ and $K''$ are \emph{Lorentz} not only when $K$ and $K'$ are inertial, but also when $K$ and $K'$ are two uniformly accelerated frames, provided that each frame experiences the same acceleration.

In section \ref{spacetimetrans}, we used the Weak Hypothesis of Locality to obtain local spacetime transformations (formula (\ref{xy})) from a uniformly accelerated frame $K'$ to an inertial frame $K$. These transformations extend the Lorentz transformations. We have also computed (equation (\ref{metricaty})) the metric at an arbitrary point of $K'$. The metric depends only on position, and not on \emph{time}.

In the process of solving the examples of section \ref{examples}, we used the ``drift frame." What is the physical meaning of this frame? What is the physical significance of the drift velocity?

In an upcoming paper, we will obtain velocity and acceleration transformations from $K'$ to $K$. We will also derive the general formula for the time dilation between clocks located at different positions in $K'$. It turns out that this time dilation depends on the state of the clock, that is, on its position and velocity.

In computing the spacetime transformations from a uniformly accelerated frame, we used the Weak Hypothesis of Locality, which is an extension of Einstein's Clock Hypothesis. Not all physicists agree with this
hypothesis. L. Brillouin (\cite{Brill}, p.66) wrote that ``we do not know and should not
guess what may happen to an accelerated clock."  It is shown in \cite{FG10} that if the Clock Hypothesis does \emph{not} hold, then there is a universal limitation $a_{max}$ on the magnitude of the 3D acceleration $\mathbf{g}$.
In \cite{FS1}, we showed that the 3D acceleration must be replaced by an antisymmetric tensor $A$ in order to achieve covariance. Thus, we expect that if the Clock Hypothesis is not valid, then the maximal acceleration will put a bound on the admissible acceleration tensors. In this case, the set of admissible acceleration tensors will form a \emph{bounded symmetric domain} known as a $JC^*$-triple (see \cite{F04}).

In \cite{F11Ann}, the first author shows how to modify the 3D Relativistic Dynamics Equation to a 3D Extended Relativistic Dynamics Equation in order to preserve the bound on accelerations. In order to make this extended equation covariant, one needs to apply a similar procedure to that used in \cite{FS1} to make the 3D Relativistic Dynamics Equation covariant. In this way, we hope to obtain the spacetime transformations from a uniformly accelerated frame in case the Clock Hypothesis is not valid.
\vskip0.4cm\noindent

We would like to thank B. Mashhoon, F. Hehl, Y. Itin, S. Lyle, and {\O}. Gr{\o}n for challenging remarks which have helped to clarify some of the ideas presented here.

\end{document}